\documentclass[reprint,nofootinbib,superscriptaddress,amsmath,amssymb,aps]{revtex4-1}
\usepackage{graphicx}
\usepackage{rotating}
\usepackage{dcolumn}
\usepackage{bm}
\usepackage{color}
\usepackage{float}
\usepackage{booktabs}
\usepackage{stfloats}
\usepackage{subfigure}
\usepackage{multirow}
\usepackage{array}
\usepackage{mathptmx, textcomp}
\usepackage[latin1]{inputenc}
\usepackage{braket}
\usepackage[colorlinks,linkcolor=magenta,anchorcolor=cyan,citecolor=blue]{hyperref}
\usepackage[multiple]{footmisc}

\def\be{\begin{equation}}
\def\ee{\end{equation}}
\def\ba{\begin{eqnarray}}
\def\ea{\end{eqnarray}}

                      \def\pd {\partial}   
              \def\.{\cdot}
\def\math {\mathcal}
\begin{document}

\title{Splitting the Echoes of Black Holes in Einstein-nonlinear Electrodynamic Theories}
\author{Aofei Sang}
\email{aofeisang@mail.bnu.edu.cn}
\affiliation{College of Education for the Future, Beijing Normal University,
Zhuhai, 519087, China}
\affiliation{Department of Physics, Beijing Normal University, Beijing 100875, China}
\author{Ming Zhang}
\email{mingzhang@jxnu.edu.cn}
\affiliation{Department of Physics, Jiangxi Normal University, Nanchang 330022, China}
\author{Shao-Wen Wei}
\email{weishw@lzu.edu.cn}
\affiliation{Lanzhou Center for Theoretical Physics, Key Laboratory of Theoretical Physics of Gansu Province,
School of Physical Science and Technology, Lanzhou University, Lanzhou 730000, China}
\affiliation{
Institute of Theoretical Physics $\&$ Research Center of Gravitation, Lanzhou University, Lanzhou 730000, China}
\author{Jie Jiang}
\email{Corresponding author. jiejiang@mail.bnu.edu.cn}
\affiliation{College of Education for the Future, Beijing Normal University,
Zhuhai, 519087, China}
\date{\today}
\begin{abstract}
    Black hole echo is an important observable that can help us better understand gravitational theories. We present that the non-linear electrodynamic black holes can admit the multi-peak effective potential for the scalar perturbations, which can give rise to the echoes. After choosing suitable parameters, the effective potential can exhibit a structure with more than two peaks. Putting the initial wave packet released outside the peaks, we find that the time-domain profile of the echo will split when the peaks of the effective potential change from two to three. This is a phenomenon of black hole echo and it might be possible to determine the geometric structure of the black hole according to this phenomenon through gravitational wave detection.
\end{abstract}
\maketitle
\section{Introduction}

As the first discovery of the gravitational wave (GW) from the binary black hole merged, the study of the gravitational theory enters a new era. Generally, the GW signal from binary massive object merge can be divided into three stages \cite{LIGOScientific:2016lio,Cardoso:2016rao,Berti:2015itd}: the inspiral stage, the merger stage, and the ringdown stage. The post-Newtonian approximation can be used to describe the inspiral stage \cite{Blanchet:2013haa} and the merger stage can only be simulated numerically \cite{Sperhake:2011xk}. As for the ringdown stage, the spacetime is closed to stationary and we can use the quasinormal modes (QNMs) to reflect the ringdown phase if the object formed at the end is a black hole \cite{Berti:2009kk,Cardoso:2016rao}.

As more GW signals are detected, more intriguing phenomena are noticed. Particularly, Abedi analysed the data of GW150914, GW151226, and LVT151012 and claimed that there are echoes in these signals \cite{Abedi:2017isz,Abedi:2016hgu}. However, their analysis is still disputed \cite{LIGOScientific:2021sio}. In spite of this, many theoretical discussions about echoes arise since echoes contain important information and can help us understand the spacetime structure better.
In the beginning, it is believed that whether a final object possesses an event horizon can be determined by the ringdown signal. This perspective is proved to be incorrect since a horizonless compact object may also produce no echo and it is found that the production of the echo is associated with the light ring \cite{Cardoso:2016rao,Cardoso:2016oxy}, which means the echo signal arises only when the effective potential has at least two peaks.
Recently, echoes are found in the ringdown stage in various of horizonless exotic compact objects, such as the wormhole \cite{Bueno:2017hyj,Bronnikov:2019sbx,Churilova:2019cyt,Liu:2020qia,Ou:2021efv}. Except for the exotic compact object, the echo was also found in quantum black holes \cite{Manikandan:2021lko,Chakravarti:2021jbv,Chakravarti:2021clm}.
Moreover, it has been shown in Ref. \cite{Dong:2020odp} that the massive gravity also gives a characteristic double-peak potential and it may lead to gravitational waves echoes. Later, in Ref. \cite{Huang:2021qwe}, a classical black hole that satisfies the dominant energy condition is found to have echo signals in an Einstein non-linear electrodynamic theory. This is unusual since the effective potential for most of the classical black hole only has one peak in general. After that, echo signals are also found in the hairy black hole \cite{Guo:2022umh}. 

Most of the previous discussion about the black hole echoes focuses on the case where the effective potential has two peaks. In Ref. \cite{Li:2019kwa}, Li and Piao first considered the late-time GW ringdown waveform when the potential exhibits more than two peaks, and they observed the characteristic phenomenon of mixing echoes. In their setting, the near-horizon regime of the black hole is modeled as a multiple-barriers filter from the quantum structure, which is implemented by adding some Delta barriers manually at the near-horizon regime and the explicit dynamic remains unknown. It is natural to ask if a similar structure and phenomenon can be found in the dynamical and classical black holes.

In the present paper, we would like to consider the general Einstein non-linear electrodynamic theories. The non-linear electrodynamic theories are first proposed by Born and Infield, which is known as Born-Infeld theory \cite{Born,Borninfield}, to resolve the issue that the point charge has infinite self-energy. In the gravitational theory, the introduction of the non-linear electromagnetic field may avoid the black hole singularity \cite{Bambi:2013caa,Ayon-Beato:1998hmi,Hassaine:2007py,Hassaine:2008pw,Soleng:1995kn}. Moreover, recent study indicate that the non-linear electrodynamic theory allows many horizons \cite{Gao:2021kvr} and multi-critical points \cite{Tavakoli:2022kmo}.
Because this series of theories have high degrees of freedom, we can choose a proper coupling constant such that the effective potential of the black hole has multiple peaks. In these cases, multiple peaks can appear far away from the horizon and it is not just a correction to the near horizon regime. Then, we will calculate the time-domain profile for the scalar perturbation. We find that the three-peak profiles have obvious differences from the two-peak profile. When the number of the peak increases to three, the wave packet will split. Through this extraordinary phenomenon, we can judge how many peaks are there for a black hole potential formed after the binary black hole merge according to the GW signal.

This paper is organized as follows: In Sec. \ref{sec2}, we introduce the general Einstein non-linear electrodynamic theories. Then we give the equation of motion of a massless scalar field perturbation and propose the boundary condition. In Sec. \ref{sec3}, we introduce the numerical method to calculate the time-domain profile. Then, in Sec. \ref{sec4}, we show some representative results. Finally, we draw a conclusion and give a discussion in Sec. \ref{sec5}.

\section{Equations of motions in the Einstein-nonlinear electrodynamic theory}\label{sec2}

The general action of the Einstein gravity minimally coupled with the non-linear electromagnetic field can be written as
\be\begin{aligned}\label{action}
S=\int d^4 x\sqrt{-g}\left(R+\math{L}_\text{EM} \right)\,,
\end{aligned}\ee
where $\math{L}_\text{EM}=-\sum_{i=1}^{\infty} a_i \math{F}^i$ with $\math{F}=F_{ab}F^{ab}$ and $F_{ab}=\nabla_{a}A_{b}-\nabla_{b}A_{a}$. When $\alpha_1=1$ and $\alpha_i=0$ for any $i>1$, the theory will go back to Einstein-Maxwell theory. The variation of Eq. \eqref{action} with respect to $g_{ab}$ and $A_{a}$ gives the equation of motion
\be\begin{aligned}\label{eom1}
&G_{ab}=-2\frac{\pd\math{L}_\text{EM}}{\pd \math{F}}F_a{}^cF_{bc}+\frac{1}{2}g_{ab}\math{L}_\text{EM}\,,\\
&\nabla_a\left(\frac{\pd\math{L}_\text{EM}}{\pd \math{F}}F^{ab}\right)=0\,.
\end{aligned}\ee
With the spherically symmetric ansatz
\be\begin{aligned}\label{ansatz}
 &ds^2=-f(r)dt^2+\frac{1}{f(r)}dr^2+r^2\left(d\theta^2+\sin^2\theta d\phi^2\right)\,,\\
 &A_a=A_t(r)(dt)_a\,,\\
\end{aligned}\ee
the general asymptotically flat black hole solution \cite{Gao:2021kvr}
\be\begin{aligned}
&f(r)=1+\sum_{i=1}^{\infty}b_i r^{-i}\,,\\
&A_t(r)=\sum_{i=1}^{\infty} c_i r^{-i}\,
\end{aligned}\ee
can be found. The relationship between the black hole parameters and the coupling constant can refer to Refs. \cite{Gao:2021kvr,Giesler:2019uxc,Sago:2021gbq}.

Next, we consider a massless scalar field perturbation $\psi$ on this spacetime background.
The equation of motion of the scalar field is
\be\begin{aligned}\label{eompsi}
\nabla_a\nabla^a \psi(t,r,\theta,\phi)=0\,.
\end{aligned}\ee
Considering the spacetime is spherically symmetric, the scalar field can be separated into
\be\begin{aligned}\label{separate}
\psi(t,r,\theta,\phi)=\sum_{lm}\frac{\Phi(t,r)}{r} Y_{lm}(\theta,\phi)\,.
\end{aligned}\ee
After replacing $\psi$ in Eq. \eqref{eompsi} by Eq. \eqref{separate}, we can find the equation of motion becomes
\ba\begin{aligned}\label{eomphi}
  \frac{\partial^2\Phi}{\partial t^2}-\frac{\partial^2\Phi}{\partial r_{\ast}^2}+V(r) \Phi=0\,,
\end{aligned}\ea
where
  \ba\begin{aligned}
  V(r)=\frac{l(l+1)f(r)}{r^2}+\frac{f(r)f'(r)}{r}
  \end{aligned}\ea
and $r_{\ast}$ is the tortoise coordinate that satisfies
\be\begin{aligned}
dr_{\ast}=\frac{dr}{f(r)}\,.
\end{aligned}\ee

In the physical region, when $r\to r_h$ while $r_{\ast}\to -\infty$, $V(r)$ tends to vanish, where we use $r_h$ to represent the event horizon of the black hole. And when $r\to +\infty$ while $r_{\ast}\to +\infty$, $V(r)$ also tends to vanish. Therefore, we can find that the asymptotic solutions of Eq. \eqref{eomphi} are
\be\begin{aligned}
\Phi \sim e^{-i \omega (t\pm r_{\ast})}\,,\quad r\to r_h\,,+\infty\,.
\end{aligned}\ee
Then, since we require the scalar field is pure outgoing at infinity and pure ingoing at the event horizon, the wave function should satisfy
\be\begin{aligned}\label{bc}
   & \Phi \sim e^{-i \omega (t+r_{\ast})}\,,\quad r\to r_h\,,\\
   & \Phi \sim e^{-i \omega (t-r_{\ast})}\,,\quad r\to+\infty\,.
\end{aligned}\ee

The echo will occur when the effective potential $V(r)$ possesses at least two peaks. For the black hole in the Einstein-nonlinear electrodynamic theory, if we choose the black hole parameters properly, we can construct the effective potential with multiple peaks.
In this paper, we would like to consider the $l=2$ case since $l=2$ is relevant to the GW observation \cite{Guo:2022umh}.

\begin{figure}[h]
    \centering
    \subfigure[The effective potential as a function of $r_{\ast}$. The parameters are $M=2.245$,$\,b_1=5.213,\,b_5=-3.811,\,b_9=3.247,\,b_{13}=-1.158$ for the $1$st panel, $M=2.280$,\,$b_1=5.398$,\,$b_5=-4.617$,\,$b_9=4.660$,\,$b_{13}=-1.880$ for the $2$nd panel, $M=2.282$,\,$b_1=5.404$,\,$b_5=-4.639$,\,$b_9=4.653$,\,$b_{13}=-1.855$ for the $3$rd panel, $M=2.282$,\,$b_1=5.405$,\,$b_5=-4.641$,\,$b_9=4.653$,\,$b_{13}=-1.853$ for the $4$th panel, and $M=2.2817$, $b_1=5.405$, $b_5=-4.642$, $b_9=4.652$, $b_{13}=-1.852$ for the $5$th panel.]{
    \includegraphics[width=0.5\textwidth]{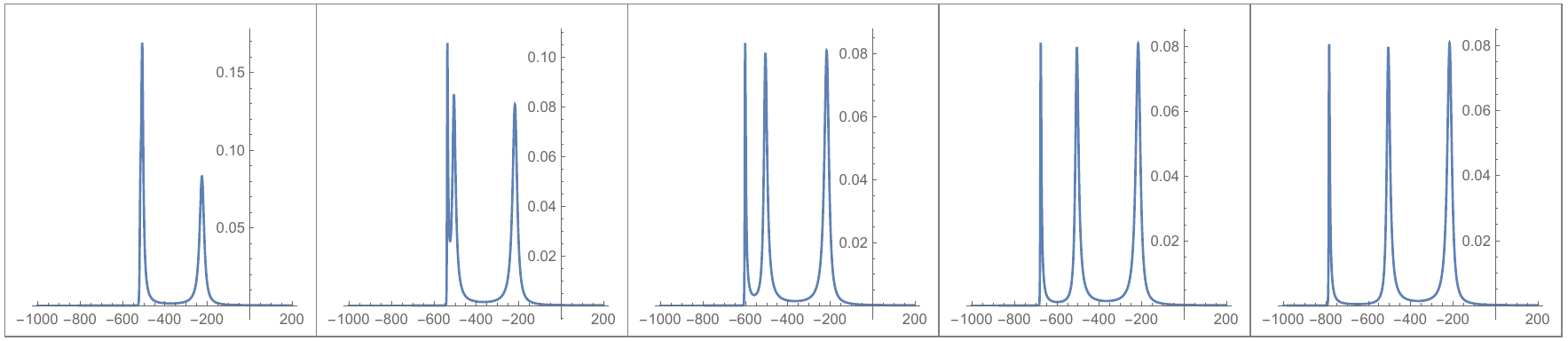}
    \label{fig11}
    }
    \\
    \subfigure[The time-domain profile corresponding to the above effective potential.]{
    \includegraphics[width=0.5\textwidth]{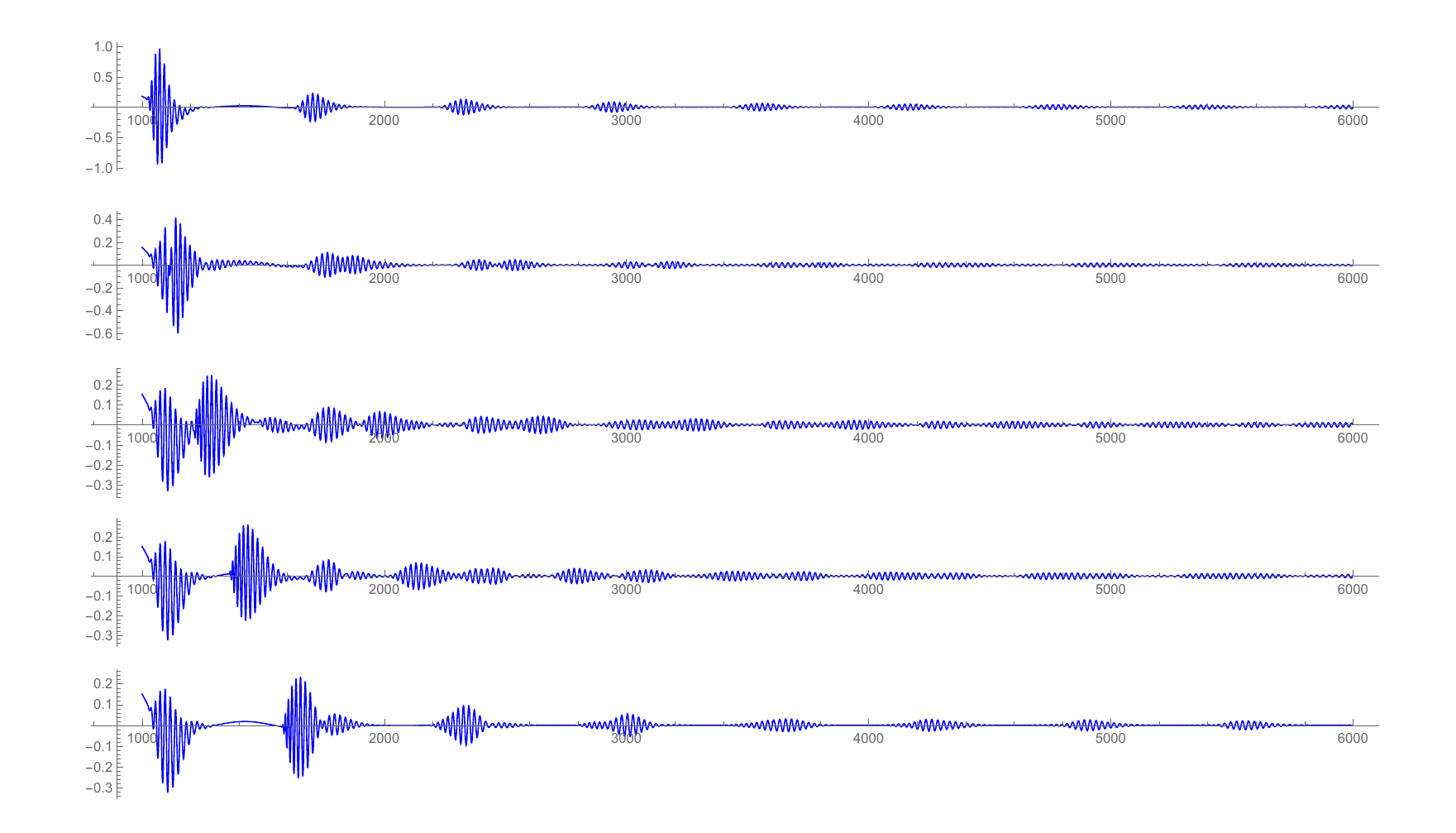}
    \label{fig12}
    }
    \caption{The top half of the figure shows a series of the effective potential, whose number of the peak changes from two to three. And the distance between the two peaks on the left increase gradually. The lower part of this figure shows the time-domain profile corresponding to the above effective potential.}
    \label{fig1}
\end{figure}

\begin{figure}[h]
    \centering
    \subfigure[The effective potential as a function of $r_{\ast}$. The parameters are $M=2.128,\,b_1=6.320,\,b_5=-10.629,\,b_9=13.679,\,b_{13}=-6.114$ for the $1$st panel, $M=2.133,\,b_1=5.009,\,b_5=-4.780,\,b_9=5.299,\,b_{13}=-2.261$ for the $2$nd panel, $M=2.277,\,b_1=5.413,\,b_5=-4.743,\,b_9=4.810,\,b_{13}=-1.927$ for the $3$rd panel, $M=2.281,\,b_1=5.406,\,b_5=-4.659,\,b_9=4.678,\,b_{13}=-1.864$ for the $4$th panel, and $M=2.282$, $b_1=5.405$, $b_5=-4.642$, $b_9=4.652$, $b_{13}=-1.852$ for the $5$th panel.]{
    \includegraphics[width=0.5\textwidth]{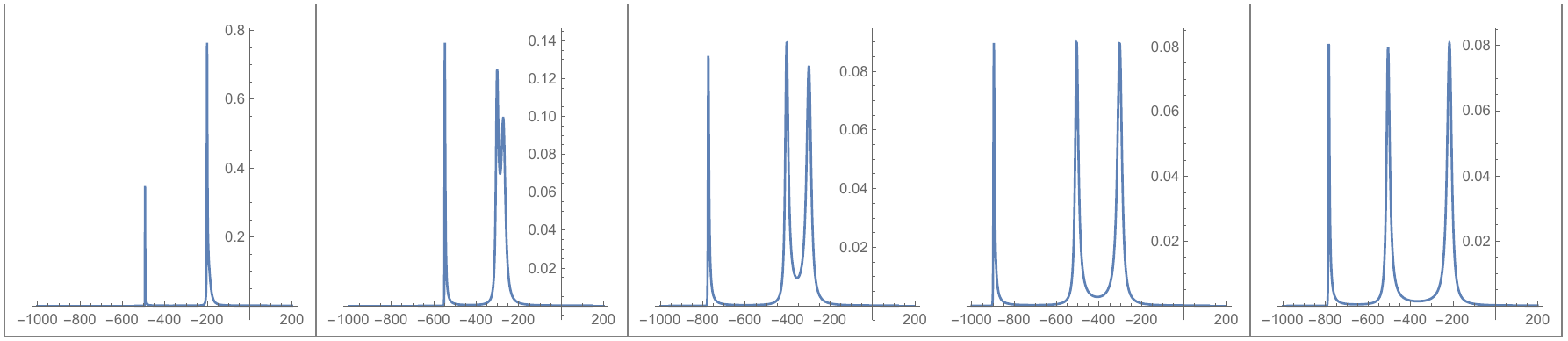}
    \label{fig21}
    }
    \\
    \subfigure[The time-domain profile corresponding to the above effective potential.]{
    \includegraphics[width=0.5\textwidth]{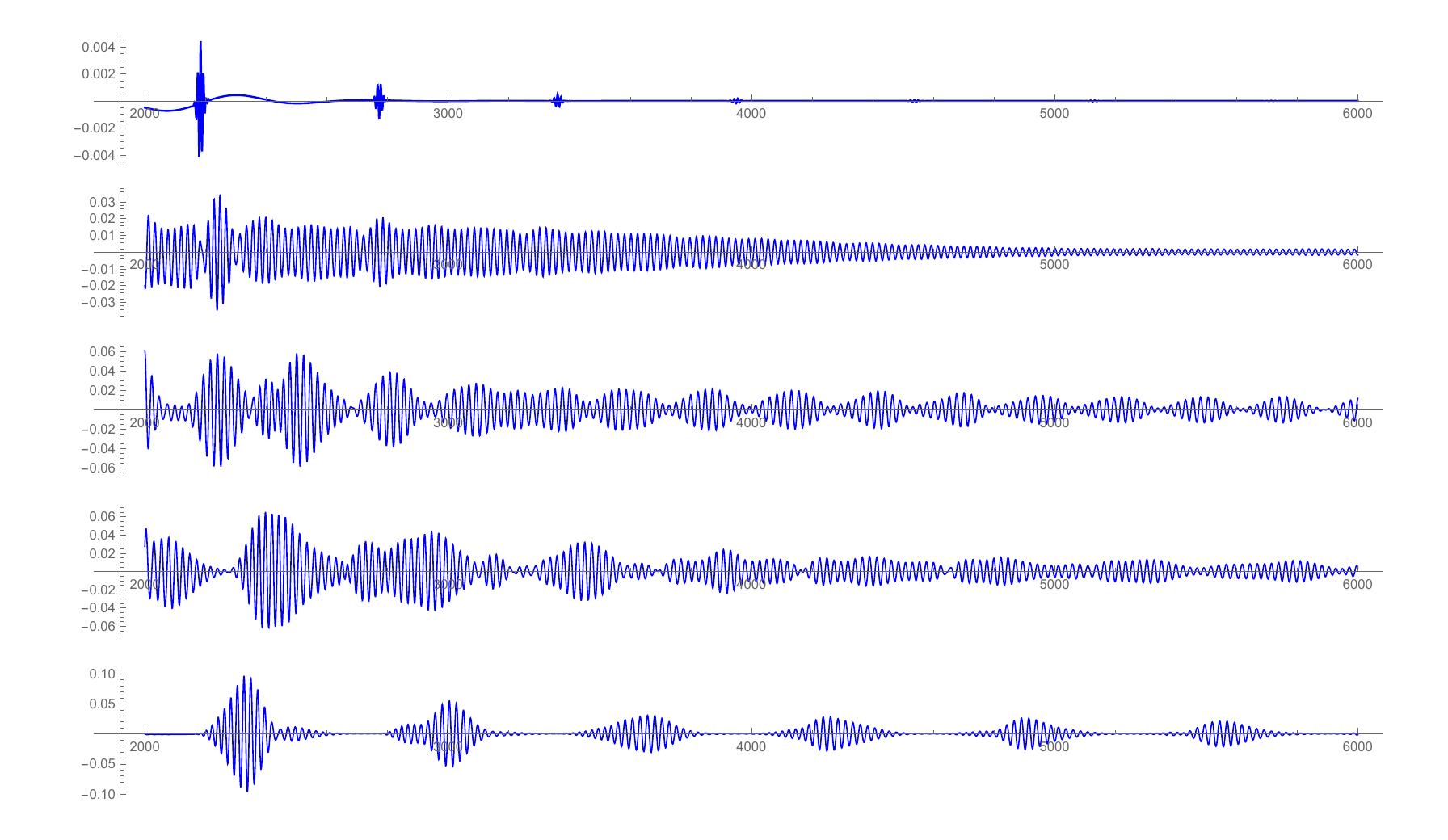}
    \label{fig22}
    }
    \caption{The top half of the figure shows a series of the effective potential, whose number of the peak changes from two to three. And the distance between the two peaks on the right increase gradually. The lower part of this figure shows the time-domain profile corresponding to the above effective potential.}
    \label{fig2}
\end{figure}

\section{Numerical method}\label{sec3}

In this section, we would like to introduce the finite difference method used to calculate the wave function in this paper and introduce how to extract the quasinormal mode from the wave function.

First, we divide the coordinate into a series of the grids. Each grid point can be represented by $(i \Delta t, j \Delta r_{\ast})$. Then, the differential equation can be cast into lots of algebraic equations:
\be\begin{aligned}
\Phi&((i+1)\Delta t,j \Delta r_{\ast})=-\Phi((i-1)\Delta t,j\Delta r_{\ast})\\
&+\Phi(i\Delta t,i\Delta r_{\ast})\left(2-2\frac{\Delta t^2}{\Delta r_{\ast}^2}-\Delta t^2 V_{\ast}(j\Delta r_{\ast})\right)\\
&+\frac{\Delta t^2}{\Delta r_{\ast}^2}\Big (\Phi\big (i\Delta t,(j+1)\Delta r_{\ast}\big )+\Phi\big (i\Delta t,(j-1)\Delta r_{\ast}\big )\Big )\,,
\end{aligned}\ee
where $V_{\ast}(r_{\ast})=V(r(r_{\ast}))$.
In this paper, we take $\Delta t=1/8$, and we choose $\Delta r_{\ast}=2 \Delta t=1/4$ considering the Neumann stability condition \cite{Zhu:2014sya}. we can solve Eq. \eqref{eomphi} numerically and we can obtain a time-domain profile after giving the initial condition
\be\begin{aligned}
\Phi(0,r_{\ast})=e^{-\frac{(r_{\ast}-a)^2}{2}}\,\quad\text{and} \quad \Phi(t<0,r_{\ast})=0\,.
\end{aligned}\ee
According to Ref. \cite{Huang:2021qwe}, the echo will become more distinct when the initial wave packet is outside the effective potential well. Therefore, we would like to set $a$, the center of the wave packet, to be zero and outside the potential well for convenience.

Next, we would like to use the Prony method to extract the quasinormal mode at late time. We take the time-domain profile from $t=t_0$ to $t=t_0+N h$, where $N$ is an integer and $h$ is the distance between each point. Here, $t_0\,,\,N$ and $h$ can be chosen freely. The profile at a certain $r_{\ast}$ can be expanded as
\be\begin{aligned}
\Phi(t)=\sum_{j=1}^{p} \tilde{C}_j e^{- i \omega_j t}\,
\end{aligned}\ee
with $p=[N/2]$, where $[x]$ denotes the integer part of $x$.

For any point we choose, this formula establish, i.e.
\be\begin{aligned}\label{prony1}
x_n=\sum_{j=1}^{p}C_j z_j^n\,,
\end{aligned}\ee
where $x_n=\Phi(t_0+n h)$, $z_j=e^{i \omega_j h}$ and $C_j=\tilde{C_j}e^{- i \omega t_0}$. Then we introduce a function
\be\begin{aligned}\label{A}
A(z)=\prod_{i=1}^p(z-z_i)=\sum_{i=0}^p \alpha_i z^i\,.
\end{aligned}\ee
It is easy to find that $A(z_i)=0$ for any integer $i$ from $1$ to $p$. Then, with an easy calculation, we can find
\be\begin{aligned}\label{prony2}
\sum_{i=0}^{p}\alpha_i x_{i+j}&=\sum_{i=0}^{p}\alpha_i\sum_{k=1}^{p}C_k z_k^{i+j}\\
&=\sum_{k=1}^{p}C_k z_k^j \sum_{i=0}^p \alpha_i z_k^i\\
&=\sum_{k=1}^{p}C_k z_k^j A(z_k)=0\,.
\end{aligned}\ee
According to Eq. \eqref{A}, we have $\alpha_p=1$. Therefore, Eq. \eqref{prony2} becomes
\be\begin{aligned}
\sum_{i=0}^{p-1}\alpha_i x_{i+j}=-x_{p+j}\,.
\end{aligned}\ee
Taking $j$ from $1$ to $p$, we will obtain $p$ equations and can easily work out $\alpha_i$. Then, we can get $z_i$ and then $\omega_i$ by finding the root of Eq. \eqref{A}. Finally, the coefficients $C_i$ can be found using Eq. \eqref{prony1}.

\begin{figure}[h]
    \centering
    \subfigure[The effective potential as a function of $r_{\ast}$. The parameters are $M=2.246,\,b_1=5.212,\,b_5=-3.803,\,b_9=3.237,\,b_{13}=-1.154$ for the $1$st panel, $M=2.282,\,b_1=5.402,\,b_5=-4.625,\,b_9=4.637,\,b_{13}=-1.851$ for the $2$nd panel, $M=2.282,\,b_1=5.404,\,b_5=-4.631,\,b_9=4.636,\,b_{13}=-1.844$ for the $3$rd panel, $M=2.282,\,b_1=5.404,\,b_5=-4.631,\,b_9=4.635,\,b_{13}=-1.843$ for the $4$th panel, and $M=2.282$, $b_1=5.404$, $b_5=-4.631$, $b_9=4.635$, $b_{13}=-1.843$ for the $5$th panel.]{
    \includegraphics[width=0.5\textwidth]{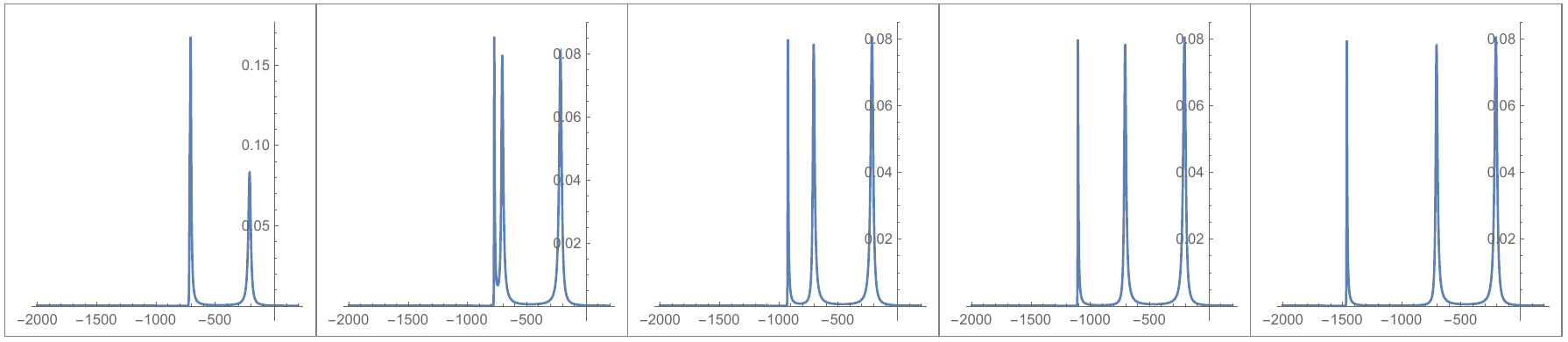}
    \label{fig31}
    }
    \\
    \subfigure[The time-domain profile corresponding to the above effective potential.]{
    \includegraphics[width=0.5\textwidth]{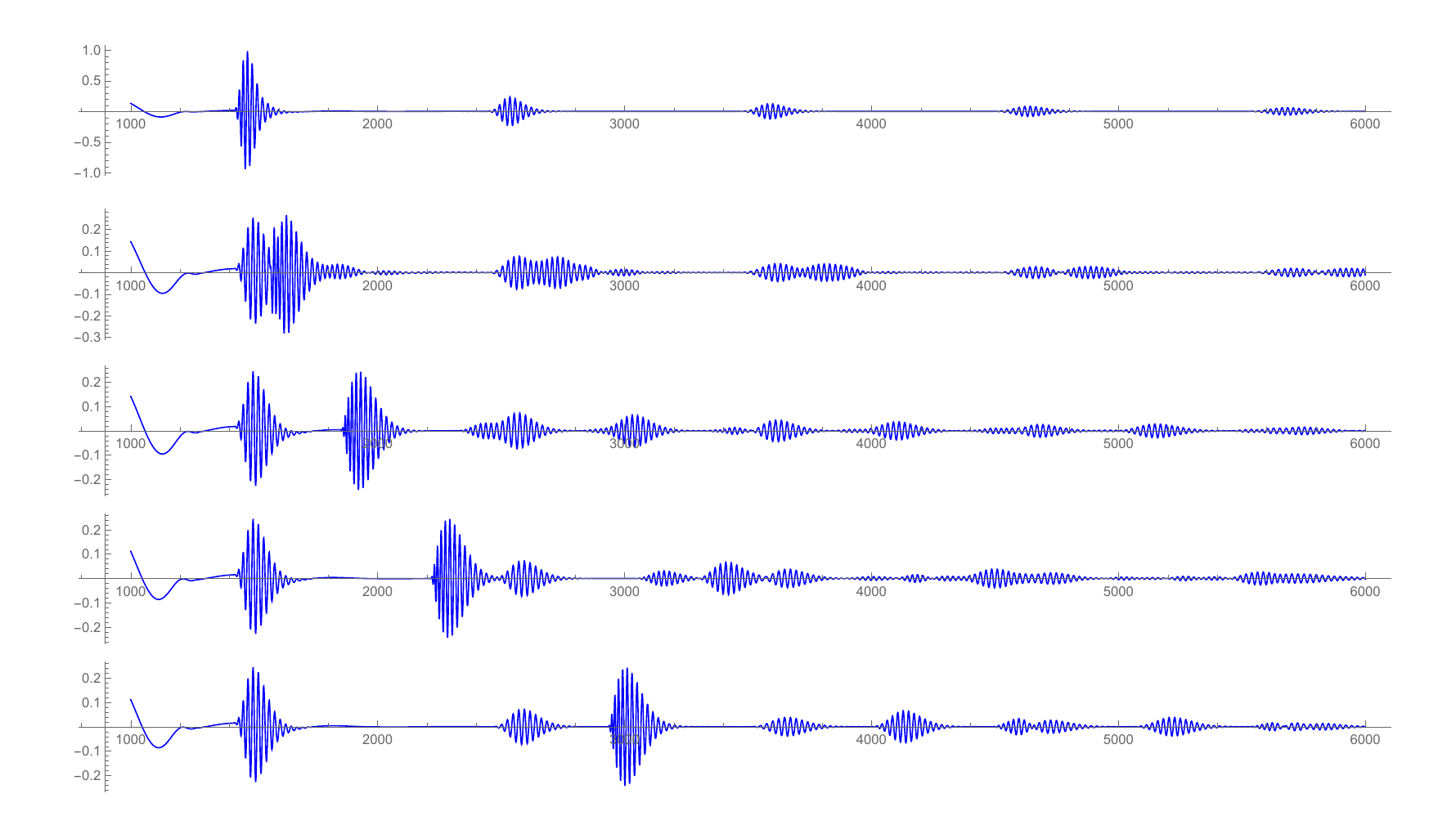}
    \label{fig32}
    }
    \caption{The top half of the figure shows a series of the effective potential, whose number of the peak changes from two to three and distance between the two peaks on the right is larger than Fig. \ref{fig1}. The distance between the two peaks on the left increase gradually. The lower part of this figure shows the time-domain profile corresponding to the above effective potential.}
    \label{fig3}
\end{figure}

\begin{figure}[h]
    \centering
    \subfigure[The effective potential as a function of $r_{\ast}$. The parameters are $M=2.128,\,b_1=6.320,\,b_5=-10.629,\,b_9=13.679,\,b_{13}=-6.114$ for the $1$st panel, $M=2.251,\,b_1=5.468,\,b_5=-5.295,\,b_9=5.663,\,b_{13}=-2.333$ for the $2$nd panel, $M=2.272,\,b_1=5.422,\,b_5=-4.838,\,b_9=4.957,\,b_{13}=-1.997$ for the $3$rd panel, $M=2.282,\,b_1=5.405,\,b_5=-4.646,\,b_9=4.659,\,b_{13}=-1.855$ for the $4$th panel, $M=2.282$, $b_1=5.404$, $b_5=-4.631$, $b_9=4.635$, $b_{13}=-1.843$ for the $5$th panel, and $M=2.283$, $b_1=5.404$, $b_5=-4.627$, $b_9=4.629$, $b_{13}=-1.840$ for the $6$th panel.]{
    \includegraphics[width=0.5\textwidth]{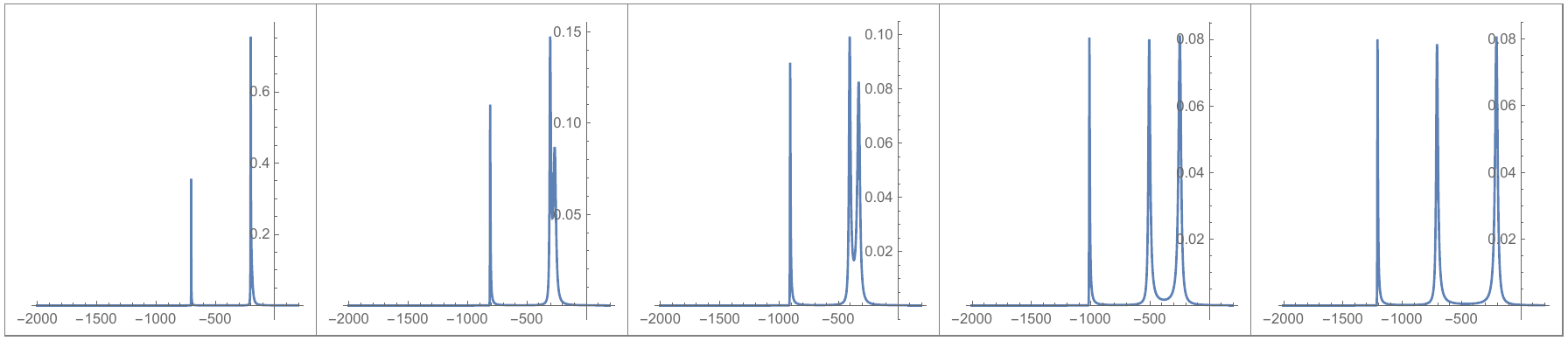}
    \label{fig41}
    }
    \\
    \subfigure[The time-domain profile corresponding to the above effective potential.]{
    \includegraphics[width=0.5\textwidth]{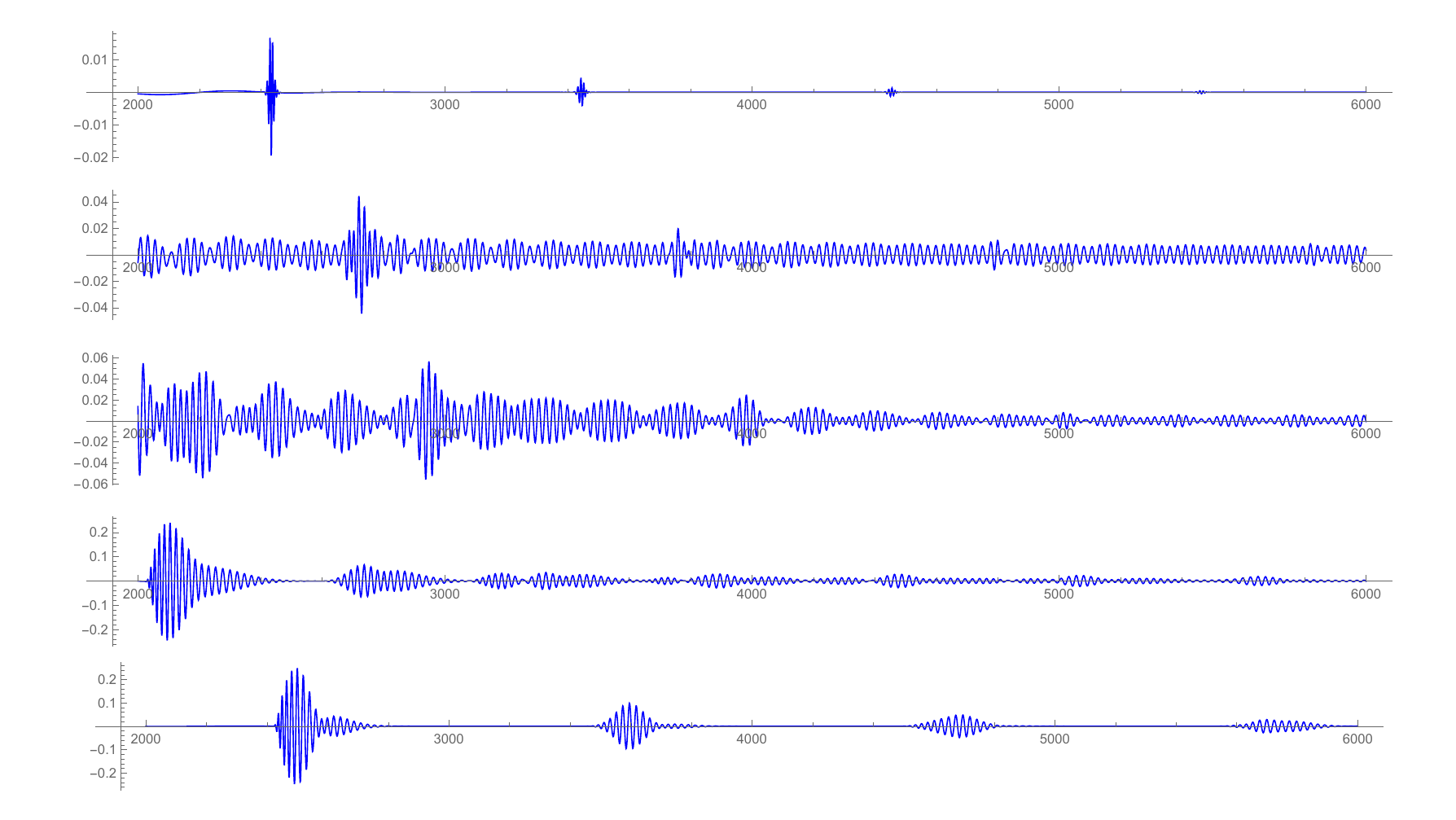}
    \label{fig42}
    }
    \caption{The top half of the figure shows a series of the effective potential, whose number of the peak changes from two to three and distance between the two peaks on the left is larger than Fig. \ref{fig3}. And the distance between the two peaks on the right increase gradually. The lower part of this figure shows the time-domain profile corresponding to the above effective potential.}
    \label{fig4}
\end{figure}

\begin{figure}[h]
    \centering
    \subfigure[The effective potential as a function of $r_{\ast}$. The parameters are $M=2.004,\,b_1=4.451,\,b_5=-3.624,\,b_9=3.703,\,b_{13}=-1.523$ for the $1$st panel, $M=2.218,\,b_1=5.194,\,b_5=-4.459,\,b_9=4.518,\,b_{13}=-1.816$ for the $2$nd panel, $M=2.280,\,b_1=5.408,\,b_5=-4.677,\,b_9=4.708,\,b_{13}=-1.879$ for the $3$rd panel, $M=2.282,\,b_1=5.405,\,b_5=-4.642,\,b_9=4.652,\,b_{13}=-1.852$ for the $4$th panel, and $M=2.282$, $b_1=5.404$, $b_5=-4.631$, $b_9=4.635$, $b_{13}=-1.843$ for the $5$th panel.]{
    \includegraphics[width=0.5\textwidth]{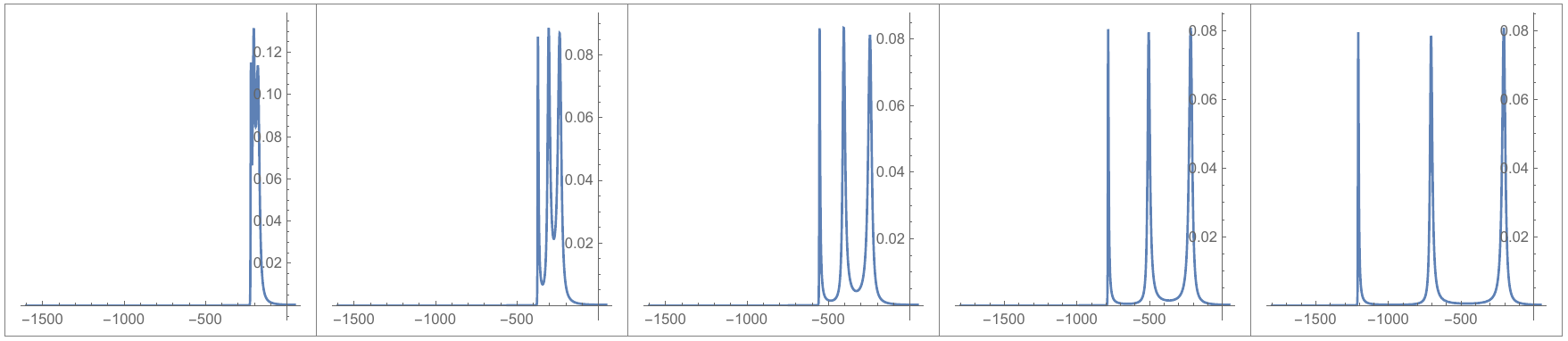}
    \label{fig51}
    }
    \\
    \subfigure[The time-domain profile corresponding to the above effective potential.]{
    \includegraphics[width=0.5\textwidth]{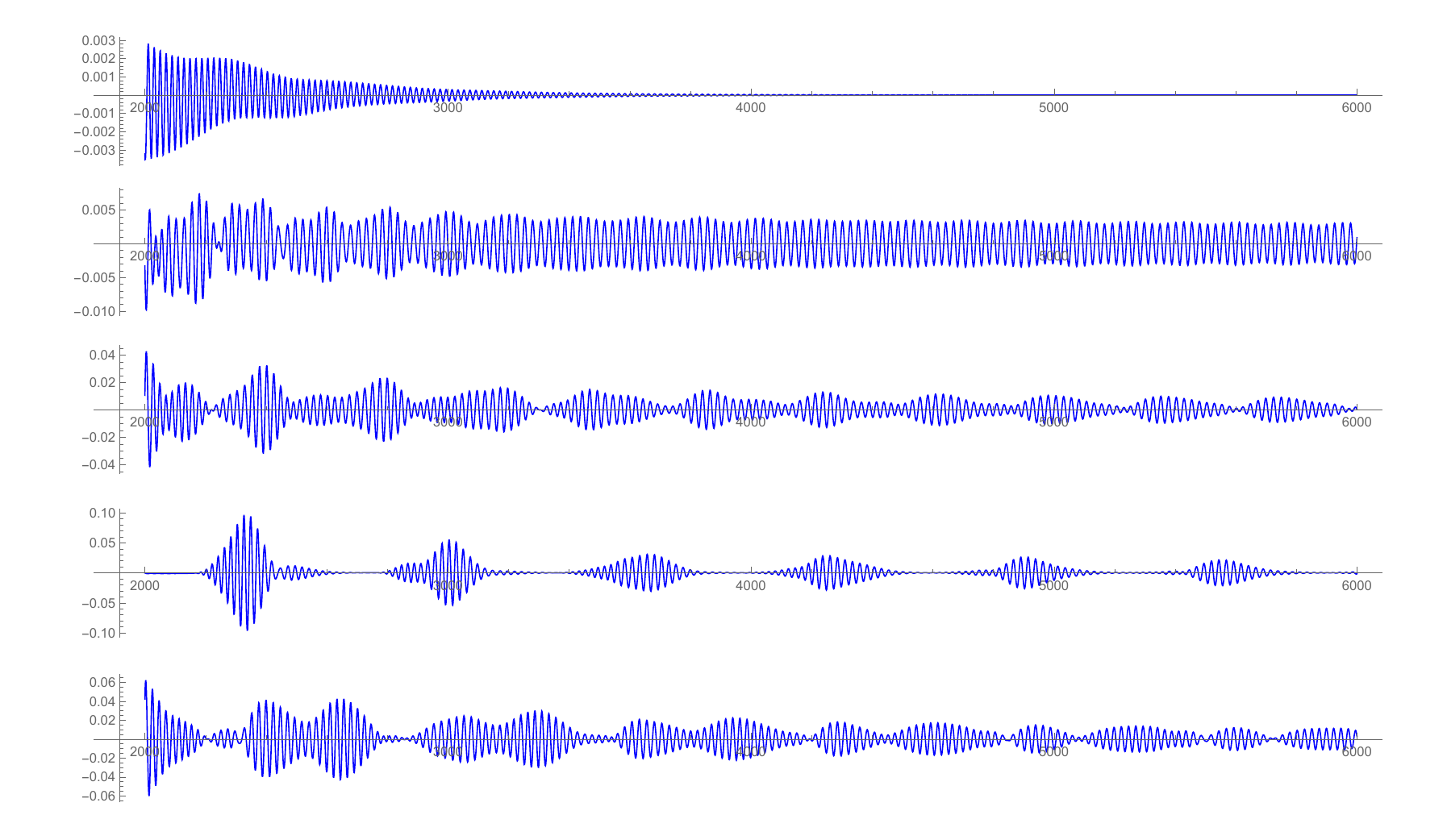}
    \label{fig52}
    }
    \caption{The top half of the figure shows a series of the effective potential, which has three peaks. And the distance between each peak increase gradually. The lower part of this figure shows the time-domain profile corresponding to the above effective potential.}
    \label{fig5}
\end{figure}

\begin{figure}[h]
    \centering
    \subfigure[The effective potential as a function of $r_{\ast}$. The parameters are $M=2.776,\,b_1=8.015,\,b_5=-15.842,\,b_9=42.025,\,b_{13}=-64.192,\, b_{17}=49.584,\,b_{21}=-15.038$ for the $1$st panel, $M=2.774,\,b_1=8.018,\,b_5=-15.930,\,b_9=42.388,\,b_{13}=-64.855,\, b_{17}=50.145,\,b_{21}=-15.218$ for the $2$nd panel, $M=2.771,\,b_1=8.021,\,b_5=-16.063,\,b_9=42.937,\,b_{13}=-65.858,\, b_{17}=50.995,\,b_{21}=-15.490$ for the $3$rd panel, and $M=2.762,\,b_1=8.035,\,b_5=-16.518,\,b_9=44.811,\,b_{13}=-69.275,\, b_{17}=53.889,\,b_{21}=-16.416$ for the $4$th panel.]{
    \includegraphics[width=0.5\textwidth]{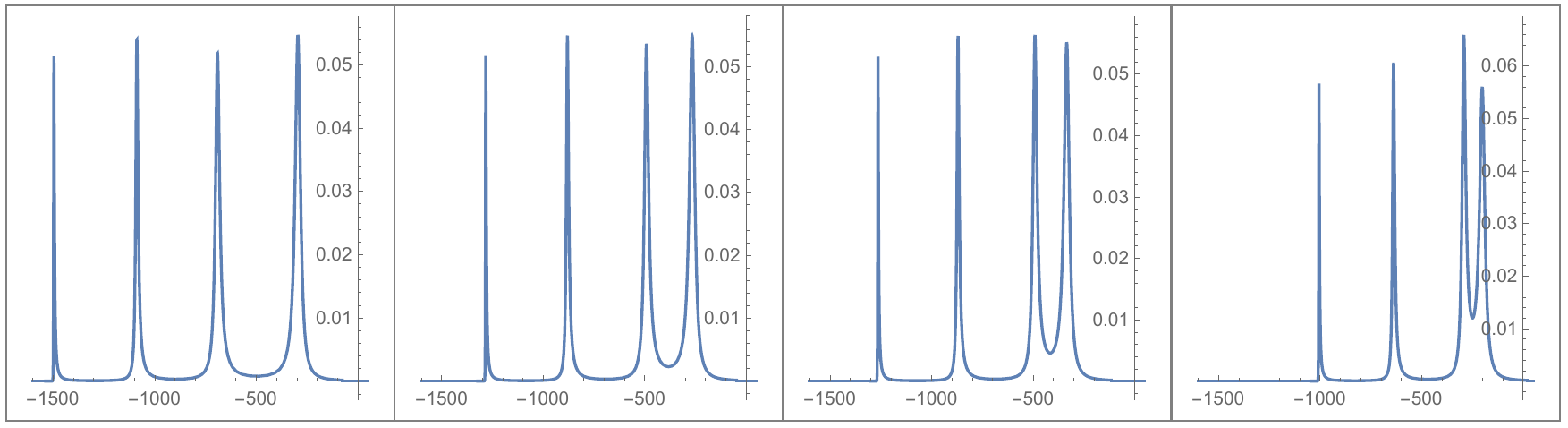}
    \label{fig61}
    }
    \\
    \subfigure[The time-domain profile corresponding to the above effective potential.]{
    \includegraphics[width=0.5\textwidth]{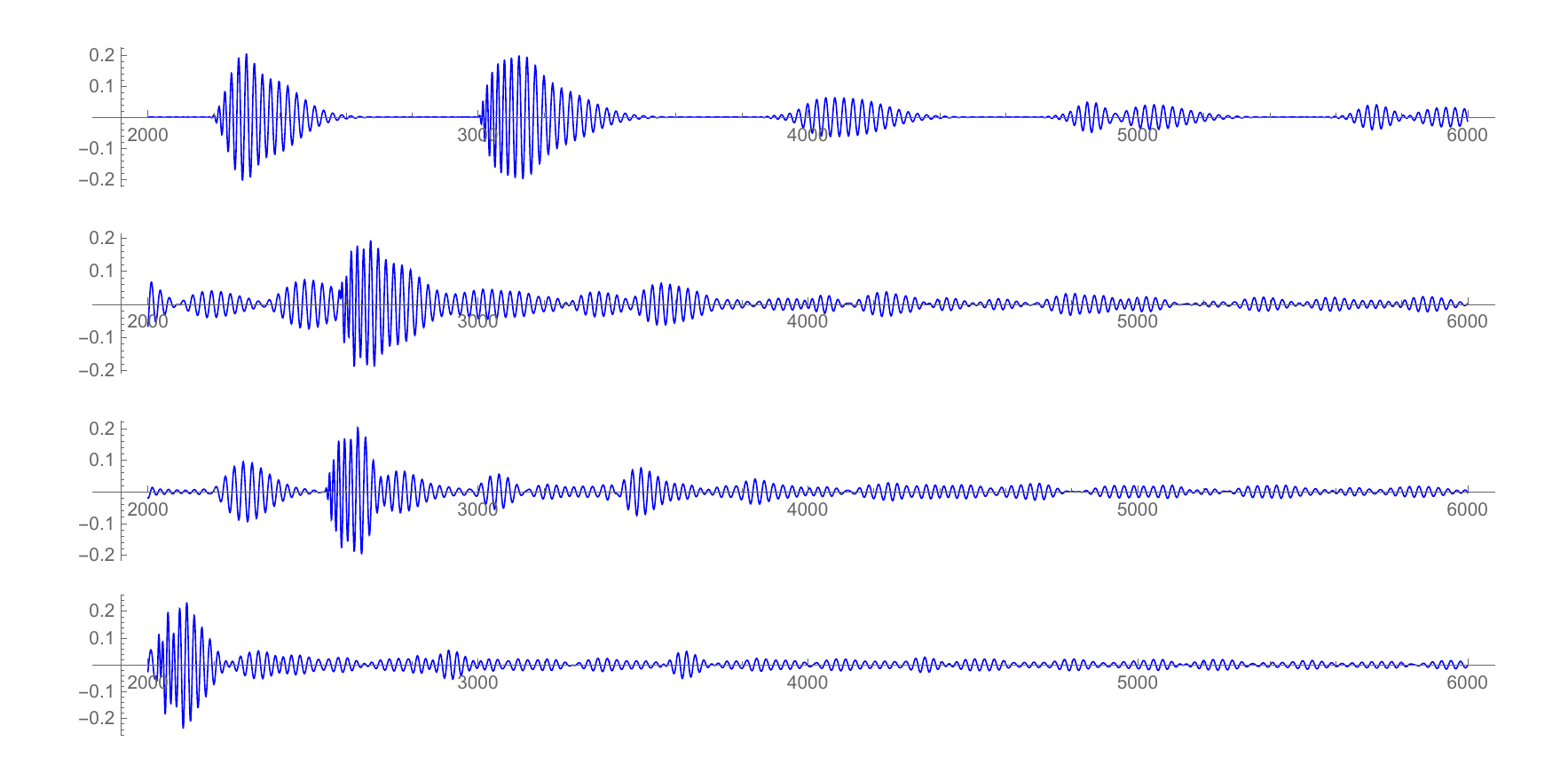}
    \label{fig62}
    }
    \caption{The top half of the figure shows a series of the effective potential, which has three peaks. And the distance between each peak increase gradually. The lower part of this figure shows the time-domain profile corresponding to the above effective potential.}
    \label{fig6}
\end{figure}

\begin{table}[h]
    \label{tab1}
    \begin{center}
        \resizebox{\linewidth}{!}{
        \begin{tabular}{|c|c|c|}
            \hline
             $n$ & $C_i$ & $\omega_i$\\
            \hline
            0 & $1.20504016\times 10^{-7}\pm 7.8226531\times 10^{-8} i$ & $\pm 0.31962997-0.00233465 i$ \\
            \hline
        \end{tabular}}
        \resizebox{\linewidth}{!}{
        \begin{tabular}{|c|c|c|}
            \hline
             $n$ & $C_i$ & $\omega_i$\\
            \hline
            0 & $-8.1623224\times 10^{-7}\pm 1.5137792\times 10^{-6} i$ & $\mp0.16641503 - 2.345211449\times 10^{-8} i$\\
            1 & $-0.0000901394624 \pm 0.000035939263 i$ & $\pm 0.20424968  -2.044901312\times 10^{-6} i$\\
            2 & $0.00053558844 \pm 0.00135318057 i$ & $\pm 0.2398756249 - 0.000092236224 i$\\
            3 & $3.84742823\times 10^{-6} \pm 3.66567045\times 10^{-6} i$ & $\mp 0.269828305 - 0.0014504918 i$\\
            $\vdots$ & $\cdots$ & $\cdots$ \\
           \hline
        \end{tabular}}
        \resizebox{\linewidth}{!}{
        \begin{tabular}{|c|c|c|}
            \hline
             $n$ & $C_i$ & $\omega_i$\\
            \hline
            0 & $-4.5043462\times 10^{-9}\pm 3.7065901\times 10^{-9} i$ & $\pm 0.1124426094364851 -2.0346910513932473\times 10^{-10} i$\\
            1& $-2.032899073\times 10^{-7}\pm 4.851510085\times 10^{-7} i$ & $\pm 0.1490018856-2.334110289\times 10^{-8} i$\\
            2& $-3.5763000969\times 10^{-6}\pm 5.426665687\times 10^{-7} i$ & $\mp 0.167012399-7.7177825\times 10^{-8} i$\\
            3& $-9.01957411\times 10^{-9}\pm 6.22909582 \times 10^{-8} i$ & $\pm 0.13080686-2.66078244\times 10^{-7} i$\\
            4& $0.0000188878\pm 9.993436960216902\times 10^{-6} i$ & $\mp 0.184818-3.668992215998481\times 10^{-7} i$\\
            5& $-0.000108466\pm 0.000024693 i$ & $\mp 0.20239-1.8448632603310853\times 10^{-6} i$\\
            6& $-0.000407296\pm 0.000295869 i$ & $\pm 0.219687-9.115572894194049\times 10^{-6} i$\\
            7& $-0.00166814\pm 0.000555234 i$ & $\mp 0.236671-0.0000442548 i$\\
            8& $0.00228217 \pm 0.000846305 i$ & $\mp 0.253401-0.00022198 i$\\
            9& $1.7825026\times 10^{-8}\pm 1.696437\times 10^-8 i$ & $\pm 0.210844-0.000613594 i$\\
            10& $1.2596547\times 10^{-7}\pm 4.984605\times 10^{-8} i$ & $\pm 0.230318-0.000879513 i$\\
            11& $0.0000855974 \pm 0.0000156702 i$ & $\pm 0.270536-0.000985614 i$\\
            12& $-0.0000209027\pm 0.0000410629 i$ & $\mp 0.26738-0.000999219 i$\\
            13& $-0.0000581573\pm 0.0000130955 i$ & $\mp 0.283665-0.00122495 i$\\
            $\vdots$ & $\cdots$ & $\cdots$ \\
            \hline
        \end{tabular}}
        \caption{The quasinormal frequency $\omega_i$ and its corresponding coefficients $C_i$ of the first three profiles in Fig.\ref{fig5}. And the first column of the table is the overtone number $n$.}
    \end{center}
\end{table}
\section{Numerical results}\label{sec4}

In this section, we will show some typical results. When the nonlinear electromagnetic field is considered, we can find the effective potential of some black holes has three peaks. When the number of the peak of the effective potential and the distance between the peaks changes, the time-domain profile will show different shapes.

First, in Fig. \ref{fig1} and Fig. \ref{fig3}, we show the time-domain profile corresponding to the various effective potential. In Fig.\ref{fig1}, the number of the peak of the effective potential split from two to three, and the distance between the two peaks on the left increases. We can find that the time-domain profile also split as the potential changes. Except for the main wave packet which already exists in the two-peak case, another wave packet appears near the main wave packet.
When the peak separation on the left and right tend to be the same, the distance between two adjacent wave packets is almost the same as the case where the potential has only two peaks on the right. The difference is that the time-domain profile decays more slowly for the three peaks case.
In Fig. \ref{fig3}, we also study another set of effective potentials whose distance between the two peaks on the right is larger than Fig. \ref{fig1}. In the profile produced by this set of the effective potential, we also find the split of the wave packet. Besides, by comparing Fig. \ref{fig1} with Fig. \ref{fig3}, we can find that the distance between the neighboring wave packets increase when the distance between the two peaks on the right side increases.

Then, in Fig. \ref{fig2} and Fig. \ref{fig4}, we choose a series of effective potentials which has an increased length of the potential well on the right side and present the relevant time-domain profile below the potential. We find that the distance between the wave packet becomes very small when the left peak, the peak which is close to the center of the initial wave packet, begins to split. As the increase of the length of the potential well on the right, the distance between the wave packets increases, and the wave function decays more slowly.
In this case, the split of the main wave packet is not obvious, but we can find that it still exists if we refer to the fourth panel in Fig. \ref{fig2}.
And when the length of the two well becomes almost the same, the distance between the wave packet is equal to the case where the potential has only two peaks. Moreover, in this case, the width of each wave packet will become larger.
Comparing Fig. \ref{fig2} and Fig. \ref{fig4}, we can find that the space between the main wave packet, the wave packet with large amplitude, increase when the distance between the two peaks on the left increase, which is similar to the conclusion drawn according to Fig. \ref{fig1} and Fig. \ref{fig3}.

Next, in Fig. \ref{fig5}, we show the time-domain profiles and their effective potential. In this set of effective potential, the space between the three peaks increases gradually. When the space is pretty small, the behavior of the time-domain profile is consistent with the one-peak potential case, where the profile decays all the time and no echo occurs. When the space becomes larger but still is small, the boundary of each wave packet is not obvious. Then, as the space increase, the wave packet becomes more distinct.

Analyzing all the above effective potential and time-domain profiles, we can find that the distance between the neighboring main wave packet is associated with the distance between the two peaks on the right side. The distance between the split wave packet and the main wave packet is associated with the distance between the two peaks on the left side. When the length of the left well and the length of the right well is approached, the split signal tends to overlap with the next main wave packet.

We also extract the frequency at the late time from the first three profiles in Fig. \ref{fig5}, which is shown in Tab. \ref{tab1}. When the distance between the three peaks is small enough, the profile is similar to the case where the effective potential has only one peak. At the late time, there is only one dominant mode, which means the wave function will always decay as is shown in Fig. \ref{fig5}. Then, as the length of the potential well increase, there will be more modes that determine the shape of the time-domain profile together. Among these modes, we can find two long-lived modes, which has a very very small imaginary part. This kind of mode decays very slowly and will last for a long time. We can reconstruct the time-domain profile using the coefficients and frequencies in Tab. \ref{tab1}, which can compare with the observational GW signal.

\section{Conclusion and discussion}\label{sec5}

We considered the Einstein gravity coupled with a non-linear electromagnetic field, which admits the black hole solutions with multi-peaks effective potential for the massless scalar field perturbation. For the multi-peaks cases, the inner potential peaks can give rise to the echoes of quasinormal modes. For some suitable parameters, the effective potential of the scalar perturbation has more than two peaks and it will lead to a different echo signal than the two-peak case. To be more specific, we mainly considered the situation where the effective potential of the black hole has three peaks and consider the process in which the number of peak changes from two to three. We also showed some profile when the effective potential has four peaks in Fig. \ref{fig6} and found that the similar splitting phenomenon. As a result, we showed that the time-domain profile of the scalar field will split for multi-peak cases. This means that we can recognize the number of the peak of the effective potential according to the echo signal.

There are still lots of issues that we can consider further except in the discussion presented in this paper. Firstly, we only considered the cases where the effective potential has three peaks in this article. It is also interesting to focus on the more peak cases. Secondly, to get closer to astronomical observations, it is necessary to study echoes of the gravitational and electromagnetic waves. Finally, we let the initial wave packet
released outside the peaks. However, it has been shown in Ref. \cite{Huang:2021qwe} that the echoes will different when the wave packet is released inside the peaks. Moreover, the echo signal will be different if we choose the different initial condition. Therefore, the influence of the initial wave packet is also worthy to be studied in our situation.

\section*{Acknowledgement}
We are grateful to Zhoujian Cao for useful discussion. J. J. is supported by the National Natural Science Foundation of China with Grant No. 12205014, the Guangdong Basic and Applied Research Foundation with Grant No. 217200003 and the Talents Introduction Foundation of Beijing Normal University with Grant No. 310432102. M. Z. is supported by the National Natural Science Foundation of China with Grant No. 12005080. S. W. is supported by the National Natural Science Foundation of China with Grant Nos. 12075103 and 12047501.

\end{document}